\documentclass[journal=jacsat,manuscript=article]{achemso}

\usepackage{chemformula}
\usepackage[T1]{fontenc}
\usepackage{latexsym}
\usepackage{subcaption}
\usepackage{balance}
\usepackage{caption}
\usepackage{amsmath}
\usepackage{amssymb}
\usepackage{esdiff}
\usepackage{ragged2e}
\usepackage{lineno}
\usepackage{gensymb}
\usepackage{enumerate}
\usepackage{hyperref}
\usepackage{mathrsfs}
\usepackage{bm}
\usepackage{float}
\usepackage{soul}

\author{Radhika Soni}
\email{radhikasoni@alum.iisc.ac.in}
\affiliation{\textit{Department of Instrumentation and Applied Physics, Indian Institute of Science, Bangalore, 560012, India}}
\author{Suvronil Datta}
\affiliation{\textit{Department of Instrumentation and Applied Physics, Indian Institute of Science, Bangalore, 560012, India}}
\author{Robin Bajaj}
\affiliation{\textit{Department of Physics, Indian Institute of Science, Bangalore, 560012, India}}
\author{Saisab Bhowmik}
\affiliation{\textit{Department of Instrumentation and Applied Physics, Indian Institute of Science, Bangalore, 560012, India}}
\author{Shinjan Mandal}
\affiliation{\textit{Department of Physics, Indian Institute of Science, Bangalore, 560012, India}}
\author{Baladitya Suri}
\affiliation{\textit{Department of Instrumentation and Applied Physics, Indian Institute of Science, Bangalore, 560012, India}}
\author{Kenji Watanabe}
\affiliation{\textit{Research Center for Electronic and Optical Materials, National Institute for Materials Science, 1-1 Namiki, Tsukuba 305-0044, Japan}}
\author{Takashi Taniguchi}
\affiliation{\textit{Research Center for Materials Nanoarchitectonics, National Institute for Materials Science,  1-1 Namiki, Tsukuba 305-0044, Japan}}
\author{Manish Jain}
\email{mjain@iisc.ac.in}
\affiliation{\textit{Department of Physics, Indian Institute of Science, Bangalore, 560012, India}}
\author{U. Chandni}
\email{chandniu@iisc.ac.in}
\affiliation{\textit{Department of Instrumentation and Applied Physics, Indian Institute of Science, Bangalore, 560012, India}}

\title[An \textsf{achemso} demo]
  {Enhanced Phonon-Assisted Tunneling in  Metal - Twisted Bilayer Graphene Junctions}


\begin{document}

\begin{abstract}
  We report planar tunneling spectroscopy measurements on metal-WSe$_2$-twisted bilayer graphene heterostructures across a broad range of gate and bias voltages. The observed experimental features are attributed to phonon-assisted tunneling and the significantly high density of states within the moir\'{e} bands. A notable finding is the enhanced phonon-assisted tunneling in twisted bilayer graphene compared to Bernal bilayer graphene, which arises from a more relaxed in-plane momentum matching criterion. Theoretical calculations of phonon dispersions enable us to identify low-energy phonon modes in both Bernal and twisted bilayers of graphene, thereby elucidating the underlying mechanism of tunneling. Our results establish planar tunneling as a versatile tool to further understand electron-phonon coupling in twisted van der Waals materials.
\end{abstract}


Twisted bilayer graphene (tBLG) and its analogues have emerged as powerful new platforms that host strong correlations aided by their dispersionless low-energy bands\cite{bhowmik2024emergent, andrei2020graphene}. These materials show complex phase diagrams with numerous emergent phases that can be tuned via band filling, including superconductivity \cite{cao2018unconventional, yankowitz2019tuning,Lu2019,PhysRevLett.121.217001}, correlated insulators\cite{cao2018correlated, saito2020independent, stepanov2020untying}, charge density waves \cite{bhowmik2022broken}, and ferromagnetism \cite{sharpe2019emergent, chen2020tunable,bhowmik2023spin}. It is widely regarded that the quenched kinetic energy of the moir\'{e} flat bands enhance electron-electron interactions, stabilizing these exotic phases. Notably, the underlying moir\'{e} potential also modifies the phonon spectra leading to renormalization and creation of phonon minibands \cite{gadelha2022electron, koshino2019moire, liu2022moire}. Although theoretical studies predict that the potential strongly modifies the electron-phonon coupling\cite{PhysRevLett.121.257001}, its implications on electrical and opto-electronic transport have remained less understood.  Electron-phonon scattering is also expected to have ramifications in the context of superconductivity, which is currently debated to be either governed by an all-electronic mechanism\cite{cao2020strange} or a conventional phonon-mediated one\cite{polshyn2019large, wu2019phonon}. However, in moir\'{e} systems, probing the phonons by standard electrical transport is arduous because the phonon spectra do not affect the transport coefficients directly.\\

In this work, we present planar tunneling as a novel probe to investigate moir\'{e} graphene. A variety of alternative transport techniques such as scanning tunneling microscopy/spectroscopy (STM/STS) \cite{li2010observation, yan2012angle, huang2018topologically, kerelsky2019maximized, xie2019spectroscopic, Tilak2021, Halle2018, kroger2020local, PhysRevB.95.161410, PhysRevLett.101.216803, endlich2014moire}, electronic compressibility \cite{tomarken2019electronic, xie2021fractional}, chemical potential measurements using single electron transistors \cite{zondiner2020cascade}, quantum twisting microscopy (QTM)\cite{inbar2023quantum}, and nano-SQUID on tip (SOT) \cite{tschirhart2021imaging,uri2016electrically} have provided invaluable information on the moir\'{e} band intricacies, which are otherwise inaccessible with standard magnetotransport measurements. However, planar tunneling, a bulk spectroscopy tool that can be operated over micrometer-sized junctions unlike the local scanning probes, was not attempted in moir\'{e} graphene thus far. Planar tunneling is a powerful tool that can probe geometries of Fermi surfaces\cite{eisenstein1991probing}, electron-electron interactions\cite{spielman2000resonantly}, defect spectroscopy\cite{chandni2015evidence}, as well as phonon energy scales\cite{gadelha2022electron,vdovin2016phonon}. In the last decade, planar tunneling in single and bilayer graphene-based van der Waals junctions using hexagonal boron nitride (hBN) \cite{britnell2012field, doi:10.1021/nl3002205} or tungsten diselenide (WSe$_2$) \cite{burg2017coherent, Burg18_PhysRevLett.120.177702} as barrier layers, have revealed a wealth of phenomena, including resonant tunneling \cite{Mishchenko2014}, negative differential conductance \cite{burg2017coherent}, large zero-bias conductance indicative of inter-layer excitons\cite{Burg18_PhysRevLett.120.177702}, and evidences for phonon \cite{chandni2016signatures} and defect-assisted tunneling \cite{chandni2015evidence}. These experiments establish the versatility and utility of planar tunneling in the context of two dimensional heterostructures. 

\section{Results and Discussion}

Here, we have studied large area planar tunnel junctions comprising tBLG with a twist angle  $\theta\approx2^\circ$ and a conventional metal (Au) as the two tunnel electrodes, and a thin WSe$_2$ layer as the tunnel barrier (inset of Fig.~1a). We have investigated tunneling conductance over a range of gate and source-drain biases. The tBLG tunneling data is compared with non-twisted ($\theta = 0$) Bernal bilayer graphene (BLG) tunnel junctions, which are otherwise geometrically identical to the tBLG junctions. Such a direct comparison allows us to elucidate various intricate details of the tunneling spectra and identify several factors manipulating electron tunneling in these systems. Most dramatically, our data highlights that the tunnel conductance is governed by the extent of overlap between the Fermi surfaces of tBLG or BLG with that of the metal, allowing for inelastic processes that relax the in-plane momentum conservation. In case of BLG, the disjoint Fermi surfaces lead to a strict conservation criterion, necessitating phonons with a minimum wavevector. On the other hand, the folded moir\'{e} bands in tBLG lead to a rather relaxed criterion for in-plane momentum conservation, resulting in enhanced tunneling. The observed signatures are in good agreement with the theoretically calculated phonon band structures of BLG/tBLG and corresponding phonon density of states (DOS).

The planar junctions in this work consist of BLG or tBLG as the top electrode, an intrinsic bilayer WSe$_2$ as the tunnel barrier, and a thin chromium-gold (Cr/Au) bottom electrode with a thickness of $\sim5/10$ nm (see Fig.~1a; the circuit for tunneling measurements is marked in red). The junctions were assembled on silicon dioxide/silicon (285 nm SiO$_2$/p-doped Si) substrates. A top hBN layer encapsulates the tunnel junctions and serves as the top gate dielectric, with a Cr/Au electrode serving as the top gate. Each device consists of 2-3 tunnel junctions allowing for reproducibility checks. Additional Ohmic contacts to tBLG/BLG can be used for concurrent transport measurements, as indicated by the circuit marked in black (see Fig.~1a).
To understand the basics of the tunneling spectra, a rudimentary theoretical formalism is introduced below.

As the simplest approximation, Bardeen's tunneling model \cite{Bardeen_PhysRevLett.6.57} describes the tunneling of electrons in a junction with top ($T$) and bottom ($B$) electrodes as:
\begin{equation}
I^e (V) \propto \int |M^e|^2 \delta (\bm{k}_T - \bm{k}_B) f(E) (1 - f(E - eV)) 
\rho_T(E) \rho_B (E - eV) dE
\end{equation}

where the tunnel current, {$I^{e}$} is directly proportional to the {elastic} tunneling matrix element {$M^{e}$}, the in-plane momentum conservation term $\delta(\bm{k}_T - \bm{k}_B)$, the DOS of the top and bottom electrodes \(\rho_T(E)\) and {\(\rho_B(E-eV)\)}, respectively, and the availability of states in the top and bottom electrodes \(f(E)\) and \([1-f(E-eV)]\), respectively {(See supporting information section V for more details on the expression and approximations)}. The in-plane momentum conservation term is crucial in determining the electron tunneling process in the junction. If the in-plane momentum is conserved, electrons tunnel elastically; otherwise, inelastic tunneling dominates. {Including the inelastic matrix elements\cite{PhysRevLett.101.216803,PhysRevB.95.161410,vdovin2016phonon,PhysRevB.93.060505} in the tunneling Hamiltonian to describe the phonon-assisted processes, the inelastic tunneling current, $I^{i}$ can be expressed as follows:}

\begin{equation}
I^{i} (V) \propto |M^i|^2 |g|^2 
\iint \delta (\bm{k}_T - \bm{k}_B \pm \bm{q}) \rho^{ph} (\omega) \rho_T (E) \rho_B (E \pm \omega + eV) \, d \omega dE
\end{equation}
{where \( M^i \) represents the average inelastic tunneling matrix element, $g$ is the average electron-phonon coupling matrix element, capturing the interaction strength between electrons and phonons, \( \rho^{{ph}}\left(\omega\right) \) is the phonon density of states for the top electrode (tBLG), and $\delta(\bm{k}_T - \bm{k}_B \pm \bm{q})$ is the momentum matching term.} From this expression, it is evident that the probability of an electron undergoing inelastic tunneling depends on the electron-phonon coupling strength, \( \left|g\right|^2 \), the phonon density of states, and the momentum matching condition. This expression naturally follows from Fermi’s golden rule, which describes transition rates for inelastic scattering events (See supporting information section V for more details on the expression and approximations).

Fig.~1b shows the three-dimensional view in momentum space for the BLG-WSe$_2$-Au junction. Fermi surface of the metal is approximated as a sphere centered around the ${\Gamma}$-point, while the BLG Fermi surfaces are tiny circles located at the ${K}$ and the ${K'}$ points, respectively. Evidently, the in-plane momentum matching criterion is not satisfied since the Fermi surfaces are disjoint, and therefore, elastic tunneling is prohibited. In Fig.~1b-c, we examine the possible inelastic processes resulting in tunneling between BLG and Au. It is evident that for each electron at the boundary of the metal Fermi surface to tunnel into the BLG Fermi surface centered around $K/K'$ points, phonons are needed. The out-of-plane momentum of the electron (vertical red lines in Fig.~1b-c) is not required to be conserved in presence of an external bias. The in-plane momentum (horizontal blue lines), on the other hand, has to be conserved and the difference between the final and initial in-plane momentum of the electron may be provided/taken away by a phonon. Hence, we can simplify the three dimensional view into a two-dimensional view as shown in Fig.~1d. The electrons that participate in tunneling which are at the surface of the metal Fermi sphere can be projected to a filled circle (in yellow), while the electrons in the BLG can be considered to be at the boundaries of tiny purple circles at ${K}$ and the ${K'}$ points. From Fig.~1d, it is clear that these two projections do not intersect and therefore elastic tunneling is heavily suppressed. However, for the tBLG case, the moir\'{e} Brillouin zone (BZ) is twist-angle dependent and much smaller than that of BLG, for low twist angles. Similar projections shown in Fig.~1e for tBLG-WSe$_2$-Au junctions clearly indicate a finite overlap between the Fermi surfaces, and therefore elastic tunneling cannot be neglected in tBLG junctions. With this basic understanding, we now describe the salient features seen in our tunneling spectra.

\begin{figure*}
\includegraphics[width=0.9\textwidth]{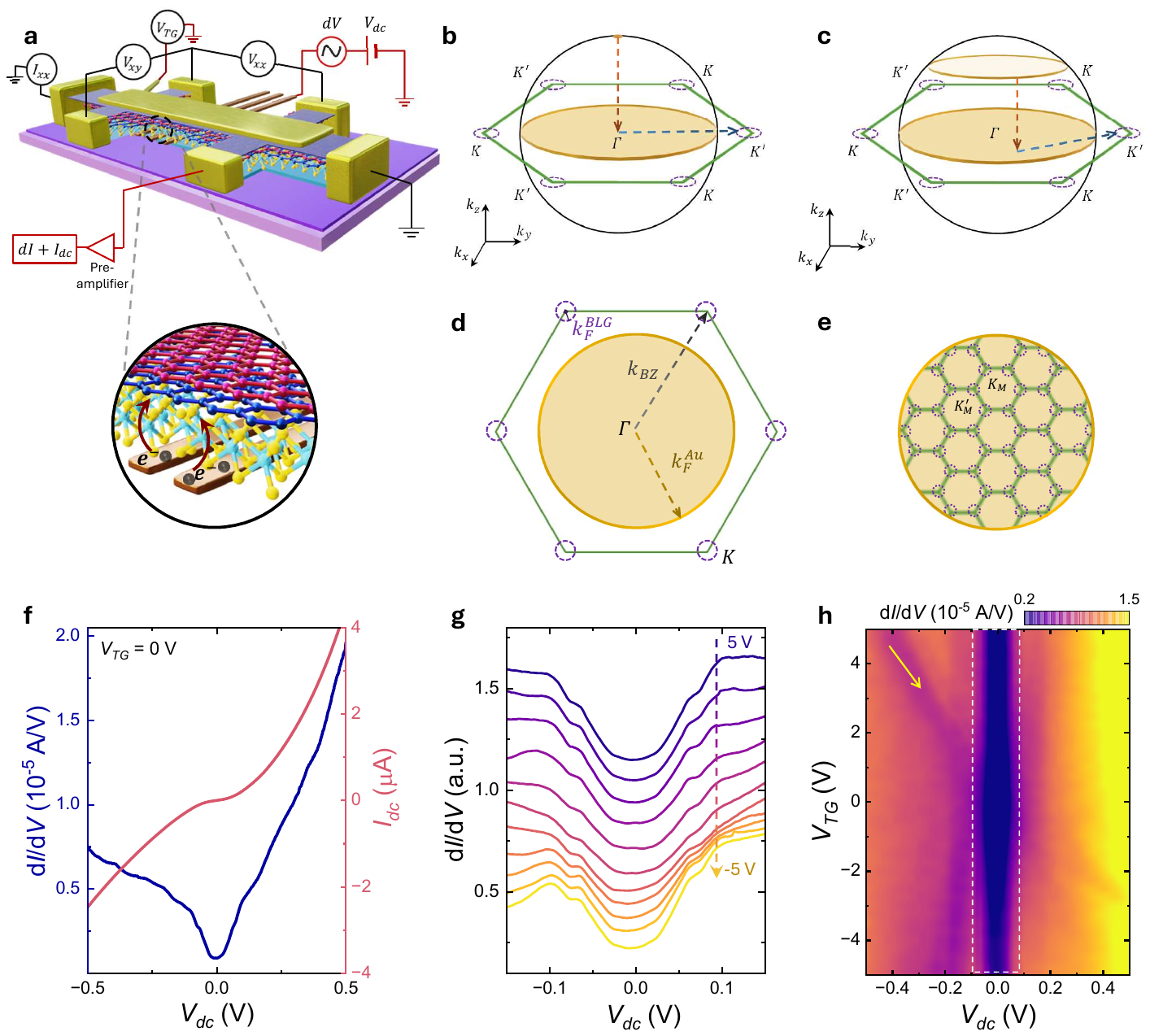}
\captionsetup{justification=raggedright,singlelinecheck=false}
\justify{\textbf{Fig.~1.~Planar tunneling in Metal-WSe$_2$-BLG/tBLG junctions}.~\textbf{a.}~Schematic of hBN-encapsulated BLG/tBLG-WSe$_2$-metal heterostructure on SiO$_2$/Si substrate.~Metal (Au) and BLG/tBLG act as the two tunnel electrodes and WSe$_2$ acts as the tunneling barrier.~The zoomed in version shows side view of the tunnel junction.~The carrier density in the system is tuned by applying top gate voltage $V_{TG}$. We perform two-probe tunneling measurements by applying an $ac$ + $dc$ bias to the metal and BLG (tBLG) electrodes, shown by the circuit in red. The circuit shown in black is used for the regular transport measurements.~\textbf{b, c.}~Schematic representations of the Fermi surfaces of metal and BLG as seen in momentum space, showing two possible inelastic tunneling scenarios, where electrons from the metal surface can use a combination of out of plane transitions and in-plane transitions to tunnel to the BLG Fermi surface concentrated around the $K/K'$ points.~\textbf{d.}~Two-dimensional view of \textbf{b, c}, which clearly shows that the two projections never intersect, resulting in suppressed elastic tunneling.~\textbf{e.}~Similar two-dimensional view of the Fermi surfaces of metal and tBLG. Unlike the BLG case, the overlapping Fermi surfaces suggest finite elastic tunneling.~\textbf{f.}~Tunnel conductance (d$I$/d$V$) and tunnel current ($I_{dc}$) as a function of $V_{dc}$ at $V_{TG}$ = 0 V in BLG device D1.~\textbf{g.}~d$I$/d$V$ as a function of $V_{dc}$ for various top gate voltages from -5 V to 5 V in steps of 1 V. Curves are offset for clarity.~The suppressed tunneling feature is present around $V_{dc}$ = 0 and the features show negligible dependence on $V_{TG}$.~\textbf{h.}~Color plot of d$I$/d$V$, plotted in a log scale as a function of $V_{TG}$ and $V_{dc}$.~The suppressed tunneling is highlighted by dashed lines and the enhancement of d$I$/d$V$ beyond $\sim 30$ mV is evident.~Tunneling is further suppressed around $V_{TG}$ = -0.63 V, which is the Dirac point of the BLG.~The yellow arrow shows the shift of the Dirac point minimum with $V_{dc}$. Measurements were performed at a temperature $T = 20$ mK.}
\end{figure*}

Tunneling current–voltage ($I_{dc}-V_{dc}$) and differential conductance d$I$/d$V$ characteristics were measured simultaneously using standard lock-in amplifier-based $ac+dc$ techniques. Fig.~1f-h presents tunneling data from a BLG/WSe\(_2\)/Au junction, denoted as D1. Fig.~1f shows tunnel conductance (d$I$/d$V$) and tunnel current (\(I_{dc}\)) as a function of dc bias, \(V_{dc}\), at zero gate voltage. A small but finite tunnel conductance is observed around zero bias (\(V_{dc} \approx 0\)), followed by an abrupt increase after $|V_{dc}|\approx30$ mV. Fig.~1g-h show d$I$/d$V$ as a function of \(V_{dc}\) for various top gate voltages, \(V_{TG}\), ranging from -5 V to $+5$ V, allowing for tuning of carrier density in BLG. The suppression of tunneling up to $ V_{dc}\approx 30$ mV is consistent across the measured range of \(V_{TG}\), resulting in a suppressed band around zero-bias in the color plot (indicated by white dashed lines). In contrast, the magnitude of d$I$/d$V$ around zero bias depends on \(V_{TG}\), and is suppressed around the Dirac point of BLG (as shown in supporting information Fig.~S1), approximately $V_{TG}$ = -0.63 V, corresponding to the minimum DOS and highest resistivity. Hence, a minimum d$I$/d$V$ around the Dirac point is not surprising. The Dirac point is also seen to shift with the application of $V_{TG}$, indicated by the yellow arrow in Fig.~1h. We also highlight that the magnitude of d$I$/d$V$ at higher $V_{dc}$ values show negligible dependence on \(V_{TG}\).\\ 
We now discuss the results of planar tunneling measurements on a Metal-WSe\(_2\)-tBLG junction, denoted as D2, where the tBLG twist angle, \(\theta\approx 2^\circ\). The twist angle is estimated using longitudinal resistivity measurements (see supporting information Fig.~S2). The deliberate choice of twist angle away from the magic angle allows us to be certain that strong correlations are negligible in these devices. {Despite the suppression of strong correlation effects such as superconductivity, ferromagnetism and correlated insulators, the tBLG system at \(\theta\approx 2^\circ\) maintains interlayer coupling, resulting in the formation of hybridized electronic bands.} Fig.~2a presents the tunnel data for one of the junctions on this device, displaying d$I$/d$V$ and \(I_{dc}\) vs. \(V_{dc}\) at  \(V_{TG}\) = 0 V. Fig.~2b and c show d$I$/d$V$ as a function of \(V_{dc}\) for various top gate voltages ranging from -5 V to 4 V. We observe both similarities and differences when compared to the data obtained for BLG- junctions (compare with Fig.~1f-h). The similarities are as follows: (1) A finite tunnel conductance is observed around \(V_{dc}\) = 0, followed by an increase beyond $|V_{dc}|\approx5-10$ mV, (2) The suppressed tunneling feature shows only slight variations with gate voltage, resulting in a dark band around $V_{dc}=0$ in the color plot shown in Fig.~2c. (3) Tunneling is maximally suppressed around \(V_{TG} \approx\) -2 V, corresponding to the Dirac point of the tBLG layer (see supporting information Fig.~S2). Next, we shall discuss the differences seen: (1) Compared with the data shown in Fig.~1h, the overall magnitude of tunnel conductance is about one order of magnitude higher for tBLG. (2) The central dark band denoting suppression of tunneling around zero bias extends to only $\approx$ 5-10 mV, as opposed to 30-40 mV in BLG.  (3) A strong enhancement of tunneling is seen across diagonal directions in Fig.~2c. This enhancement in d$I$/d$V$ originates from the Dirac point and increases diagonally in all four directions denoted by the white dotted arrows {(see supporting information section III for more quantitative details)}. Therefore, in the region of enhanced tunneling, d$I$/d$V$ shows variation with both \(V_{dc}\) and \(V_{TG}\).

To verify the reproducibility of the tunnel conductance and associated features, we conducted planar tunneling measurements on a separate junction on device D2 discussed in Fig.~2. Tunnel characteristics for this junction are presented, with different tBLG contacts in Fig.~3a and 3b (see insets).  Most importantly, we observe stark similarities such as the suppressed tunneling feature at zero bias, maximally suppressed tunneling at the Dirac point and enhancement of tunnel current in the diagonal directions. We therefore conclude that our observations are robust across junctions and are not mediated by defects or twist angle inhomogeneities.

{Previous reports~\cite{chandni2016signatures,davenport2019probing, PhysRevB.85.073405, Zhang2008, brar2007scanning, PhysRevB.102.115410, Halle2018, PhysRevB.100.075435} on metal-hBN-graphene junctions~\cite{chandni2016signatures, davenport2019probing, PhysRevB.85.073405} and STM measurements on graphene~\cite{Zhang2008, brar2007scanning, PhysRevB.102.115410, Halle2018, PhysRevB.100.075435} have shown d$I$/d$V$ spectra which are suppressed around zero bias, with enhanced tunneling seen at finite biases of $\sim 30-70$ mV (see table 1 in supporting information section IV)}. These signatures were attributed to various inelastic processes, in particular phonons in the various constituent layers, which provide the requisite in-plane momentum. {In these studies, a large zero bias gap in $dI/dV$ is consistently seen, and phonons do not contribute at low biases of $V_{dc}<30$ mV.} We observe that by simply swapping the BLG with the tBLG layer, there is an enhancement in the tunneling around zero bias, implying a relaxed conservation criteria for tBLG. {In accordance with the Eq. (1) and (2),} such an enhancement in the tunnel conductance can occur due to various reasons, including (i) better momentum matching criterion for in-plane elastic tunneling, (ii) large density of phonons at lower energy scales and (iii) enhanced electron-phonon coupling in tBLG, facilitating the tunneling process itself. {In the low-frequency range, where the phonon DOS exhibits a sharp peak associated with the layer breathing modes, both the electron-phonon coupling strength and inelastic tunneling matrix elements are not expected to vary significantly with momentum. We therefore approximate them as constants. Consequently, following Eq. (2), we primarily attribute the observed enhancement in tunneling current to the phonon density of states, \( \rho^{{ph}}\left(\omega\right) \), and momentum matching condition}. In the following, we will investigate the phonon band structures for both BLG and tBLG to better understand the experimental observations.

\begin{figure}
\includegraphics[width=1.0\textwidth]{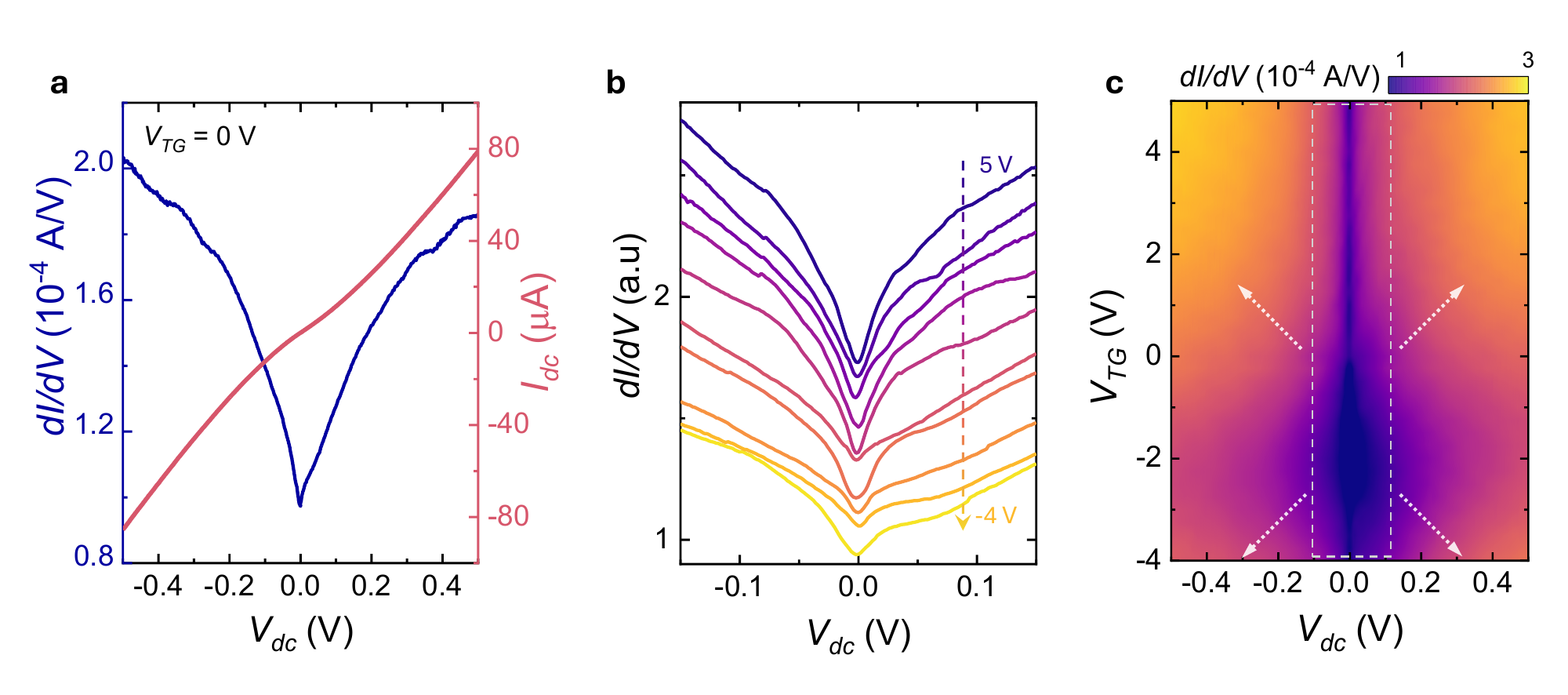}
\captionsetup{justification=raggedright,singlelinecheck=false}
\justify{\textbf{Fig.~2.~Tunnel data for Metal/WSe$_2$/tBLG junction, $\theta \approx 2^\circ$}, Device D2. ~\textbf{a.}~Tunnel conductance (d$I$/d$V$) and tunnel current ($I_{dc}$) as a function of $V_{dc}$ at $V_{TG}$ = 0.~\textbf{b.}~d$I$/d$V$ as a function of $V_{dc}$ for various top gate voltages from -5 V to 4 V in steps of 1 V. Curves are offset for clarity.~The suppressed tunneling feature is present around $V_{dc}$ = 0.~\textbf{c.}~Color plot of d$I$/d$V$, plotted in a log scale as a function of $V_{TG}$ and $V_{dc}$.~The suppressed tunneling is highlighted by dashed lines and the enhancement of d$I$/d$V$ is shown by the dashed arrows.~The enhanced tunnel conductance shows dependence on both, $V_{dc}$ and $V_{TG}$.~Tunneling is further suppressed around $V_{TG}$ = -2 V, which is the Dirac point of the tBLG. Measurements were done at $T = 4 K$.}
\end{figure}

\begin{figure}
\includegraphics[width=1\columnwidth]{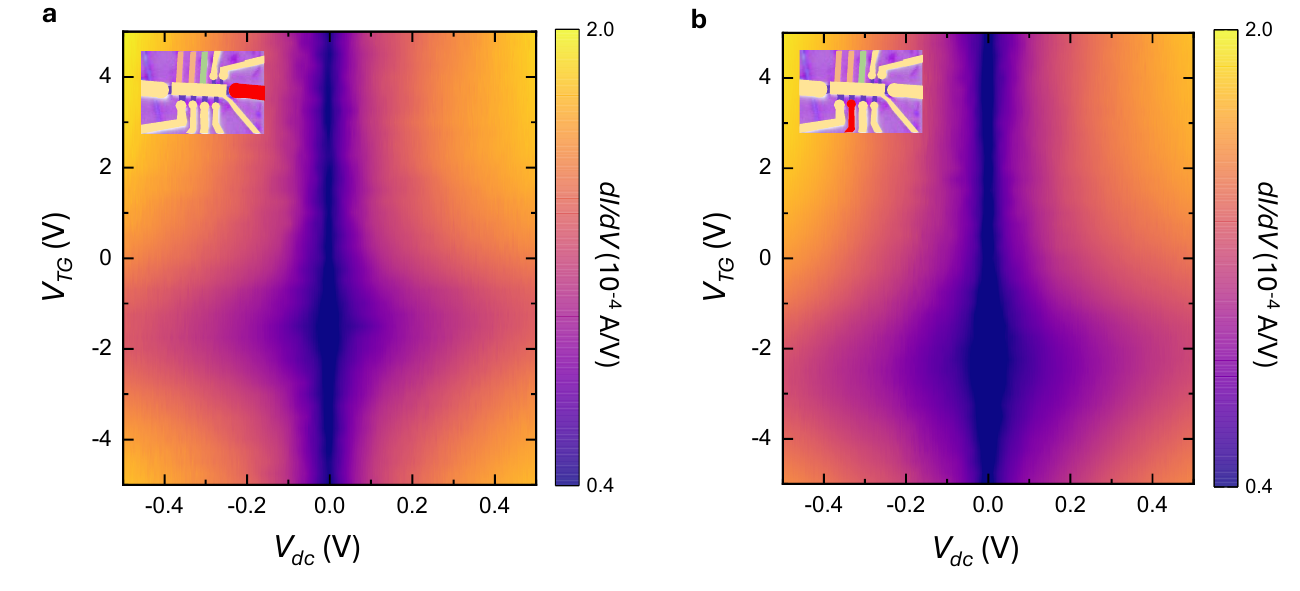}
\captionsetup{justification=raggedright,singlelinecheck=false}
\justify{~\textbf{Fig.~3.~Uniform tunnel spectra for different Metal/WSe$_2$/tBLG junctions in device D2 ($\theta \approx 2^\circ$)}.~Color plots in \textbf{a.} and \textbf{b.}  show d$I$/d$V$, plotted in a log scale as a function of \(V_{TG}\) and \(V_{dc}\) for a different junction (green colored contact) and two seperate tBLG contacts as shown (red colored contacts) in the respective inset images.} 
\end{figure}

Fig.~4a,b show the phonon band structures and density of states(DOS) for BLG and tBLG ($\theta=2^\circ$){, calculated using the PARPHOM\cite{mandal2024parphomparallelphononcalculator} package,} along the high symmetry paths of their BZ, namely ${\Gamma-M-K-\Gamma}$ for BLG and ${\Gamma_{M}-M_{M}-K_{M}-\Gamma_{M}}$ for tBLG (See Methods Section). As expected, due to band folding, both materials exhibit prominent phonon DOS peaks at similar energies, 10 meV and 50 meV (right side panels of Fig.~4a, 4b). However, experimental measurements of the tunneling conductance show significant differences between BLG and tBLG (Fig.~1g-h and 2b-c). In BLG, the tunneling current remains suppressed below a critical bias voltage of approximately $30-40$ meV, followed by an abrupt increase (see Fig.~1g-h). Conversely, tBLG exhibits an enhancement of the tunneling current around $V_{dc} = 5-10$ meV (See Fig.~2b, 2c). This implies that the tunneling in BLG is insensitive to the layer breathing mode (LBM) at 10 meV, while it plays a significant role in the enhanced tunneling in tBLG. This discrepancy arises because of an interplay between the DOS and geometric factors in the momentum space. Although both materials share the 10 meV DOS peak corresponding to the LBM, their momentum space characteristics differ significantly. {Previous studies~\cite{doi:10.1021/acs.jpclett.1c01802} of BLG on Ru substrates have shown $\Gamma-$ point ZA phonon modes at around the same energy scales ($\sim$ 14 meV). However, we speculate that for a free-standing BLG like ours, the effects of ZA modes at the $\Gamma-$ point will be minimal, as corroborated by our data. In this context, the difference seen between tBLG and BLG for the same device structure highlights that such external effects are minimal.}

To analyse this further, we look at the 2D projections of the Fermi surfaces in the momentum space. In case of BLG, as discussed in Fig~1, the Fermi surfaces are disjoint, necessitating phonons to aid in tunneling. Fig.~4c shows the shortest phonon vector that is needed to facilitate this inelastic tunneling, allowing us to write down a simple geometric relation that needs to be satisfied for in-plane momentum conservation:

\begin{equation}
    {\bm{k}}_{{BZ}}=\bm{k}_{{F}}^{{Au}}+\bm{k}_{{F}}^{{BLG}}+\Delta \bm{q}
\end{equation}
where $\bm{k}_{{BZ}}$ denotes the size of the BZ of BLG, $\bm{k}_{{F}}^{{Au}}$ and $\bm{k}_{{F}}^{{BLG}}$ denote the sizes of the Fermi surfaces for Au and BLG, respectively, and $\Delta \bm{q}$ denotes the minimum momentum mismatch to be compensated by inelastic processes. The {red dashed} circle in Fig.~4c represents the maximum in-plane momentum ($\bm{q}_{{LBM}}$) attainable by the LBM modes associated with the van Hove singularity in the DOS near 10 meV.
We assume $\bm{k}_{{BZ}}$ = $1.70 \times 10^{10}$ ${m}^{-1}$,  $\bm{k}_{{F}}^{{Au}}$= $1.20 \times 10^{10}$ ${m}^{-1}$, and $\bm{k}_{{F}}^{{BLG}}= 1.8\times10^7$ ${m}^{-1}$, for a nominal charge carrier density of $1\times10^{12}$ cm$^{-2}$, giving $\Delta \bm{q}= 0.5\times10^{10}$ ${m}^{-1}$.
The value of $\bm{q}_{{LBM}}$ is estimated to be $0.25 \times 10^{10}$ ${m}^{-1}$ from the phonon dispersion. For this mode to contribute to phonon assisted tunneling, it should be greater than $\Delta \bm{q}$ estimated above. 

Notably, $\Delta\bm{q}$ is larger than $\bm{q}_{{LBM}}$ in BLG.

In the inset to Fig. 4c, we show the zoomed in region around $K-$point. At various $k-$ points on the boundary of the Fermi surface for BLG, we represent the LBM by circles of radius $\bm{q}_{{LBM}}$, forming a halo around the Fermi surface. We observe that although the presence of LBM phonons reduces the momentum mismatch in BLG, it does not sufficiently compensate for the mismatch $\Delta q$. On the other hand, in case of tBLG, the moir\'{e} BZ is much smaller ($1.02\times 10^{9}$ ${m}^{-1}$, for $\theta = 2^\circ$), encompassing the projection of the metal Fermi surface, relaxing the criterion for  tunneling. Here, the presence of LBM further enhances the overlap between the Fermi surfaces of Au and tBLG (see Fig.~4d), facilitating enhanced tunneling mediated by the LBM at 10 meV in tBLG. Our finding highlights the crucial role of geometric compatibility between the metal Fermi surface and tBLG moir\'{e} BZ for efficient phonon-assisted tunneling.

\begin{figure*}
\includegraphics[width=1\columnwidth]{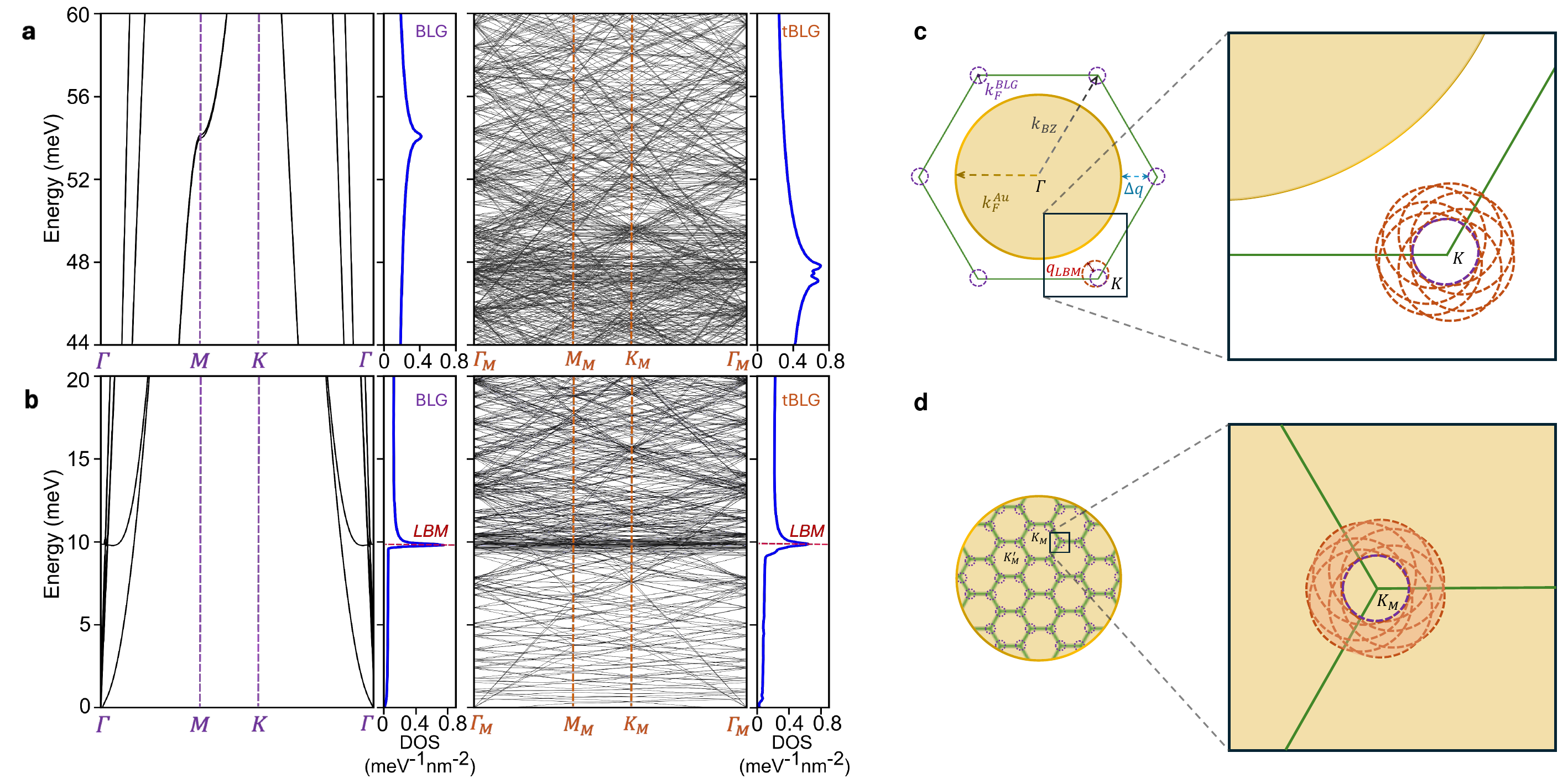}
\captionsetup{justification=raggedright,singlelinecheck=false}
\justify{\textbf{Fig.~4. Mechanism of inelastic tunneling and phonon band structures}. 
\textbf{a.} and \textbf{b.} show the phonon band structures and density of states calculated for BLG ($\theta = 0^{\circ}$) and tBLG ($\theta = 2^{\circ}$}), respectively.~\textbf{c.}~Detailed two-dimensional projection of Fermi surfaces of metal and BLG. Even though the presence of layer breathing mode (LBM) phonons decreases the momentum mismatch (depicted using a few red circles around the BLG Fermi surface), there is no overlap with the Fermi surface of metal, causing negligible elastic tunneling.~\textbf{d.}~Analogous 2-D projection of Fermi surfaces of metal and tBLG. Contrary to the BLG scenario, the LBM augments the overlapping region as shown by the shaded dashed circular region in red, resulting in enhanced tunneling.~(Moir\'{e} BZ is not to the scale)
\end{figure*}

Another key observation for tBLG systems is the dependence of d$I$/d$V$ on both \(V_{TG}\) and \(V_{dc}\), which is not observed in BLG systems (see Fig.~1h and 2c). In Fig.~2c, near the Dirac point, a region of suppressed tunneling is evident, corroborating the low DOS in tBLG. As \(V_{TG}\) and \(V_{dc}\) are adjusted, we gradually access regions of high electronic DOS in tBLG over a larger dc bias window. This scenario suggests that phonon-assisted tunneling, combined with the high availability of states in tBLG, contributes to the enhanced tunneling observed in the diagonal directions. The evidence that the band structure of the tBLG influences the tunneling process is a crucial outcome of this work, as it indicates that planar tunneling can be explored as a novel tool for investigating moiré systems.

{The tunneling data does not exhibit the intricacies of the moir\'{e} physics, owing to the diminished correlated effects around the non-magic angle. The striking resemblance of the reduced $dI/dV$ around zero bias for both BLG and tBLG, and the stability of the suppressed tunneling with applied gate voltages suggest a similar origin for both samples. Comparison with previous literature suggests that these signatures closely align with phonon-assisted events (see Table~1 in section IV in supporting information, for a compilation of various previous studies on metal-insulator-graphene based junctions). The complex nature of moir\'{e} flat bands can induce finer modifications in the tunneling response which will require more study, particularly near magic angles.}

\section{Conclusion}
In summary, we report one of the first demonstrations of planar tunneling in a moir\'{e} graphene-based junction, offering a powerful new probe to examine electron-phonon coupling and correlated phases in moir\'{e} materials. We observe that the tunneling current is indeed dominated by the extent of overlap of the Fermi surfaces of the tunnel electrodes, but the net magnitude and characteristics of the spectra are governed by the availability of phonon modes of the right wave vectors. These experiments establish twist angle as a new tunable knob that can alter the tunnel currents in van der Waals tunnel junctions. By placing our findings in a broader context, we show that while tunneling suppression with large gaps ($>30$ mV) is a consistent feature in graphitic systems with metallic electrodes, our results on twisted bilayer graphene represent the first clear deviation from this trend. By combining geometric Brillouin zone considerations with an inelastic tunneling model, we provide new insight into the tunneling mechanism and explain why low-energy phonon features have been less visible in prior studies.

\section{Methods}
 \subsection{Device fabrication}
The widely known dry-pickup transfer\cite{kim2016van} method was used to assemble the heterostructures. Polypropylene carbonate (PPC) film coated on a polydimethylsiloxane (PDMS) stamp was utilized for picking up individual layers of hBN, WSe$_2$ and BLG/tBLG. The 2D layers were obtained using the mechanical exfoliation technique. We have used the `tear and stack'\cite{kim2016van} technique to assemble the tBLG heterostructures. The sharp edge of the top hBN was used to tear the
graphene, following which one-half of the torn graphene was picked up, leaving the other half on the substrate. The sample stage was then rotated by $\theta \approx 2^\circ$, and the
second half of the graphene layer was picked up to create the tBLG. The metal tunneling electrodes were patterned on a separate 285 nm SiO$_2$/Si substrate using e-beam lithography, and  Cr/Au (5 nm/10 nm) was deposited using thermal evaporation. A thin layer of intrinsic WSe$_2$ (tunnel barrier) was transferred onto the metal tunneling electrodes. The remaining parts of the heterostructure were transferred to the prepared substrate. The devices were created by etching the heterostructure into a multi-terminal Hall bar using reactive ion etching with CHF$_3$/O$_2$. This was followed by electron-beam lithography and thermal evaporation of Ohmic edge contacts and top gate using Cr/Au (5/50nm).

\subsection{Measurement Scheme}
Electrical transport studies were conducted on a cryogen-free, pumped He$^4$ cryostat and a He$^3$-He$^4$ dilution refrigerator. A Keithley 2450 source-measure unit was used to apply gate voltages (V$_{TG}$) to the devices. A small continuous $ac$ current of around 50 nA at 13 Hz-17 Hz was supplied using an SRS Lock-in amplifier for longitudinal resistance measurements. We utilized a 1:1 isolation transformer for tunneling measurements, allowing us to distinguish between the ground of $ac$ and $dc$ signals entering the channel. The $dc$ signal (V$_{dc}$) was sourced by a 12-bit Kepco programmer (SN 488-122) with optical isolation, and an SRS 830 lock-in-amplifier was used to source the $ac$ excitation.

\subsection{Theory}
We generate the rigid tBLG structure at $\theta=2^{\circ}$ using the TWISTER code\cite{NAIK2022108184}. Structural relaxations of the atoms in the tBLG are performed using classical force fields implemented in the LAMMPS package\cite{THOMPSON2022108171}. Specifically, we use the Rebo\cite{DonaldWBrenner_2002} potential for the intralayer interactions and the DRIP\cite{PhysRevB.98.235404} potential for the interlayer interactions. The structure is relaxed upto a force tolerance of $10^{-6}$ eV/\AA. We use the PARPHOM code\cite{mandal2024parphomparallelphononcalculator,PhysRevB.110.125421} to compute the force constants for the relaxed structure using the classical interatomic potentials. The force constants are used to construct the dynamical matrix, $D(\bm{q})$, at each $\bm{q}$ point along the path ${\Gamma_{M}-M_{M}-K_{M}-\Gamma_{M}}$ of the moir\'{e} BZ. We diagonalise $D(\bm{q})$ to get the phonon spectra along the path. The DOS was calculated using a BZ integration with a dense mesh of $500\times500\times1$ $\bm{q}$-points for tBLG. The phonon spectra for BLG ($\theta = 0^{\circ}$) has been calculated using phonopy\cite{phonopy-phono3py-JPCM} and LAMMPS along the path ${\Gamma-M-K-\Gamma}$ of the unit cell BZ. The DOS was calculated on a mesh of $1000\times1000\times1$ $\bm{q}$-points for BLG.

\section*{Acknowledgements}
We thank S. Bhattacharyya for her inputs. We gratefully acknowledge the usage of the MNCF and NNFC facilities at CeNSE, IISc. U.C. acknowledges funding from SERB via SPG/2020/000164 and WEA/2021/000005. M.J. acknowledges the National Supercomputing Mission of the Department of Science and Technology, India, and Nano Mission of the Department of Science and Technology for financial support under Grants No. DST/NSM/R\&D HPC Applications$/2021/23$ and No. DST/NM/TUE/QM-10/2019 respectively. R.B. acknowledges the funding from the Prime Minister’s Research Fellowship (PMRF), MHRD. {K.W. and T.T. acknowledge support from the JSPS KAKENHI (Grant Numbers 20H00354 and 23H02052) and World Premier International Research Center Initiative (WPI),~MEXT,~Japan.

\section*{Author contributions}
R. S. and S. D. contributed equally. R.S. fabricated the devices, performed the measurements and analyzed the data. S.D. contributed to measurements and analysis. S.B. provided inputs on experiments. R.B., S.M. and M.J. planned all the theory calculations. R.B. and M.J. executed and analyzed all the theoretical calculations.~B.S. provided experimental support. ~K.W., and T.T. grew the hBN crystals.~R.S., S.D., R.B., M.J. and U.C. co-wrote the manuscript.~U.C. supervised the project.\\

\textbf{Supporting information:} Additional experimental data, analysis, comparison of tunneling in various metal-insulator-graphitic tunnel junctions, theory of elastic and inelastic tunneling, discussion on other possibilities.

\bibliography{References}

\clearpage
\newpage

\section{Supporting Information: Enhanced Phonon-Assisted Tunneling in  Metal - Twisted Bilayer Graphene Junctions}

\subsection{Electrical transport measurements}
Fig.~S1 shows the electrical characterization of device D1 with longitudinal resistance ($R_{xx}$) as a function of $V_{TG}$. The inset shows the device's optical micrograph. The Dirac point of BLG was observed to be around -0.63 $V$.\\\\
\begin{figure}[ht]
    \centering
    \includegraphics[width=0.8\linewidth]{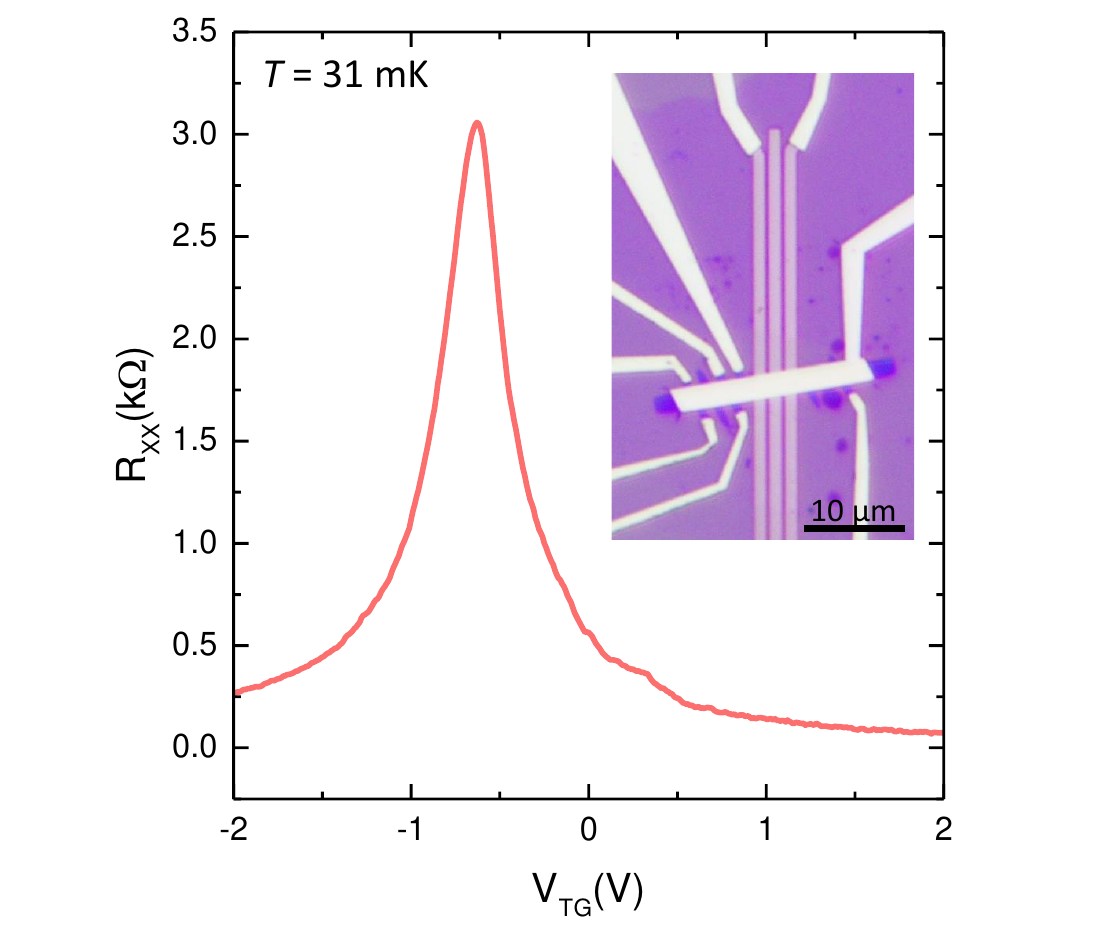}
    \captionsetup{justification=raggedright,singlelinecheck=false}
    {\textbf{Fig.~S1.~Electrical transport measurements in the device D1}. Four-probe longitudinal resistance of BLG as a function of top gate voltage (V$_{TG}$). The Dirac point of BLG is around -0.63 V. The inset shows the optical image of the final device, showing BLG patterned into a channel and electrodes in bright yellow. The light yellow color vertical electrodes are the gold tunnel electrodes with lower thickness. Scale bar is 10 $\mu$m. Measurements were done at \(T \approx\) 31 mK.}
    \end{figure}
Fig.~S2 shows electrical transport measurements on device D2. Analysis from Fig.~S2b confirms the TBG twist angle, \(\theta \approx 2^\circ\), suggesting that the two layers of SLG in tBLG exhibit weak interlayer coupling (depicted in the inset). This weak coupling results in the hybridization of tBLG bands, leading to a high DOS at lower energies. However, it can also be described by two Dirac cones from individual SLG layers displaced in reciprocal space. Consequently, as the transverse electric field magnitude increases, the weakly coupled layers are doped with an equal number of carriers of opposite signs, reducing the resistance of the tBLG. These effects are evident in our data.
\begin{figure}[ht]
    \centering
    \includegraphics[width= 1\linewidth]{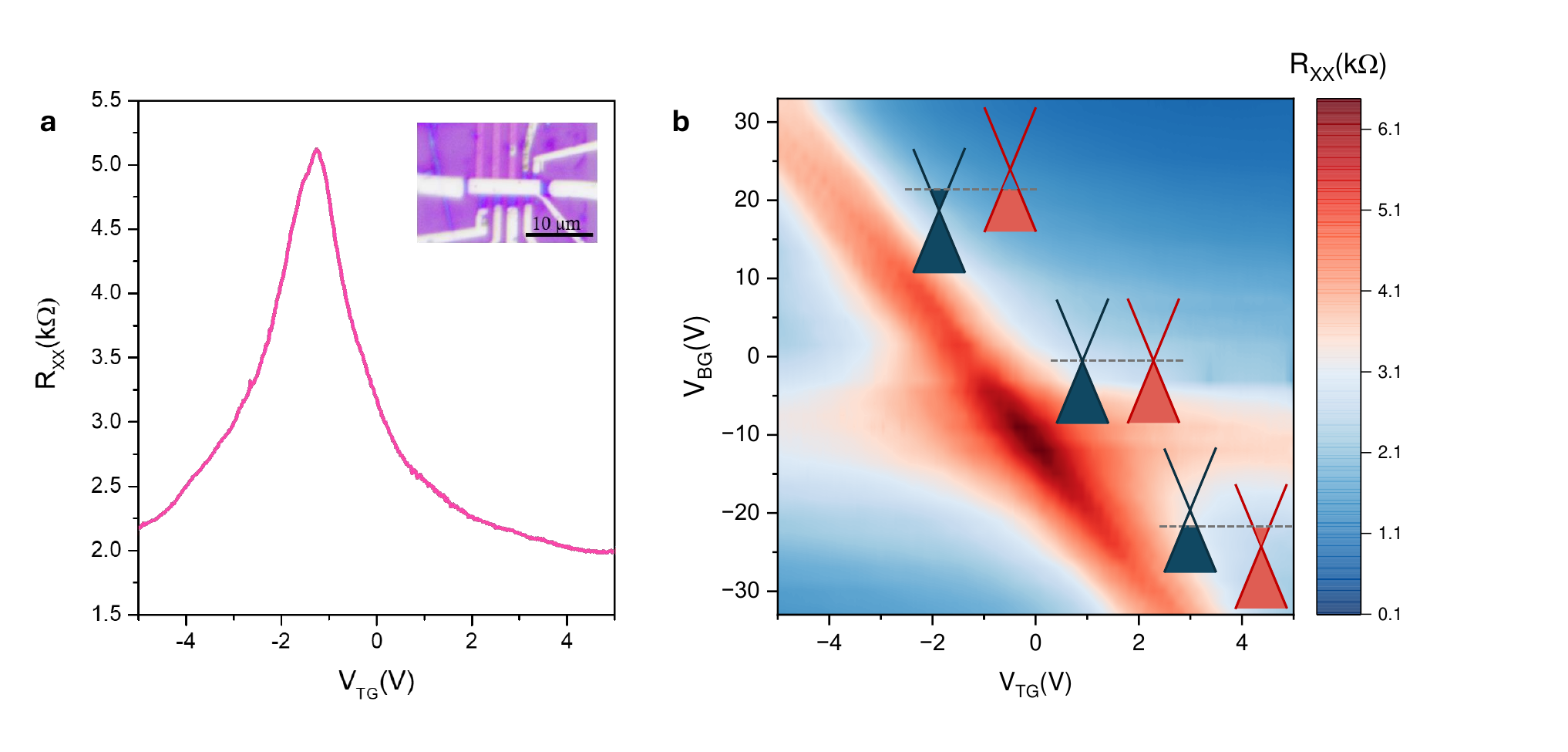}
    \captionsetup{justification=raggedright,singlelinecheck=false}
    {\textbf{Fig.~S2.~Electrical transport measurements in device D2}. \textbf{a.} Four-probe longitudinal resistance, \(R_{xx}\) of tBLG as a function of top gate voltage \(V_{TG}\). The Dirac point of tBLG is around -2 V. Inset shows the optical image of the final device showing tBLG patterned into a channel and electrodes shown in bright yellow color. The light yellow color vertical electrodes are the gold tunnel electrodes. Scale bar is 10 $\mu$m. \textbf{b.} \(R_{xx}\) as a function of top gate voltage \(V_{TG}\) and \(V_{BG}\). Contrary to BLG, the resistance was found to decrease with the application of a vertical displacement field. The inset images highlight the Dirac cones of individual graphene layers, indicating that the twist angle, \(\theta \approx 2^\circ\). Measurements were done at \(T \approx\) 4 K.}
\end{figure}

\subsection{Planar tunneling measurements on Device D3}
 We measured another device, D3, with Metal/WSe$_2$/tBLG junctions with \(\theta \approx 2^\circ\). Figs.~S3a-b show the tunnel conductance ($dI/dV$) and tunnel current ($I_{dc}$) as a function of $V_{dc}$ at $V_{TG}$ = 0 V in TBG. The tunnel electrodes are different for both plots while we kept the TBG contact to be the same. We observe an enhanced tunnel conductance beyond 10 mV as seen in device D2, confirming the repeatability of phonon-assisted tunneling. This device showed few additional features compared to D2 at higher bias voltages, which will need further investigation. Nevertheless, the behavior around zero bias was found to be consistent with D2.
 \begin{figure*}[bth!]
    \centering
    \includegraphics[width=0.9\linewidth]{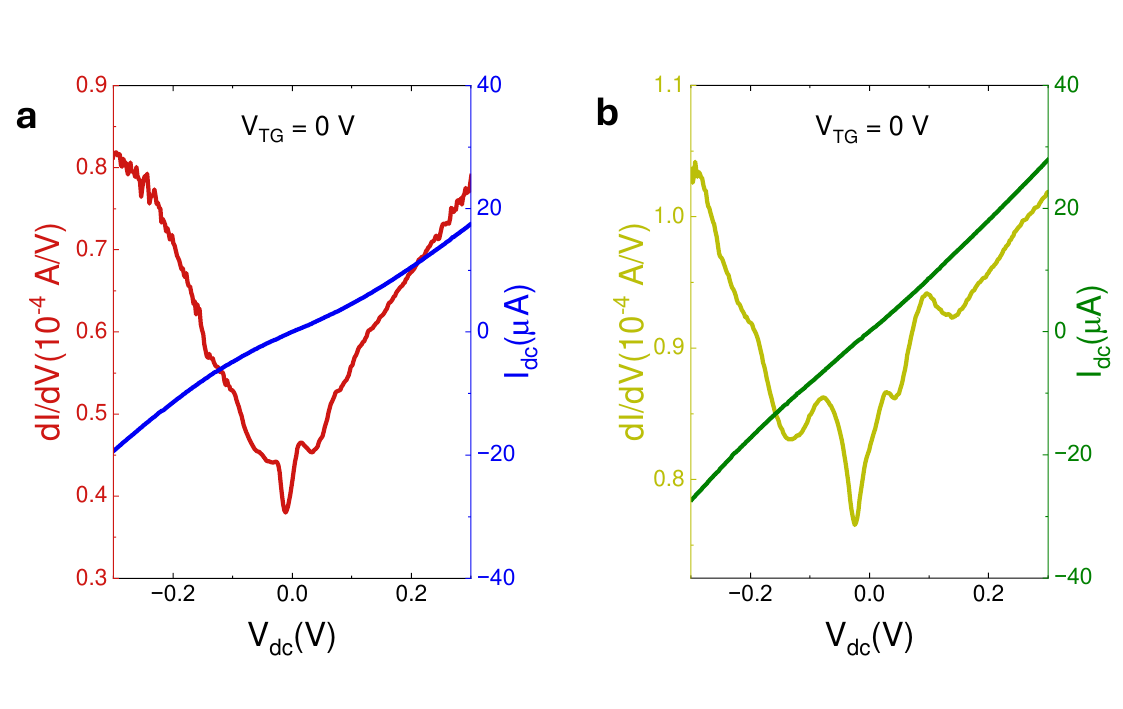}
    \captionsetup{justification=raggedright,singlelinecheck=false}
    {\textbf{Fig.~S3.~Planar tunneling measurements on Metal/WSe$_2$/tBLG junctions in device D3} ($\theta \approx 2^\circ$).~Line plots in \textbf{a.} and \textbf{b.}~show tunnel conductance ($dI/dV$) and tunnel current ($I_{dc}$) as a function of $V_{dc}$ at $V_{TG}$ = 0 V in TBG, for two different tunnel contacts keeping the TBG contact same. Measurements were done at \(T \approx\) 2 K.}
\end{figure*}

\subsection{Analysis from the $V_{DC}-V_{TG}$ colour plots}
 {
The distinct signatures of tunneling conductance measured in BLG and tBLG are shown in Figs.~1~and~2 in the main text. To further illustrate how $dI/dV$ changes in the $V_{DC}-V_{TG}$ space along the diagonal directions from the charge neutrality point, we have marked two diagonal lines in blue and maroon, for BLG and tBLG, respectively  in Fig.~S4a-b. As illustrated in Fig.~S4c, in case of BLG, the value shows marginal increase on the negative side, but remains almost constant otherwise. In contrast, for the tBLG, the enhancement of $dI/dV$ is much more robust, and an increasing trend is observed on both the positive and negative sides. The distinct nature of BLG and tBLG along the color plot is evident in this analysis. Similar results are obtained for diagonal lines throughout the color plot, indicating an enhancement in the diagonal direction.}

\begin{figure*}[bth!]
    \centering
    \includegraphics[width=0.9\linewidth]{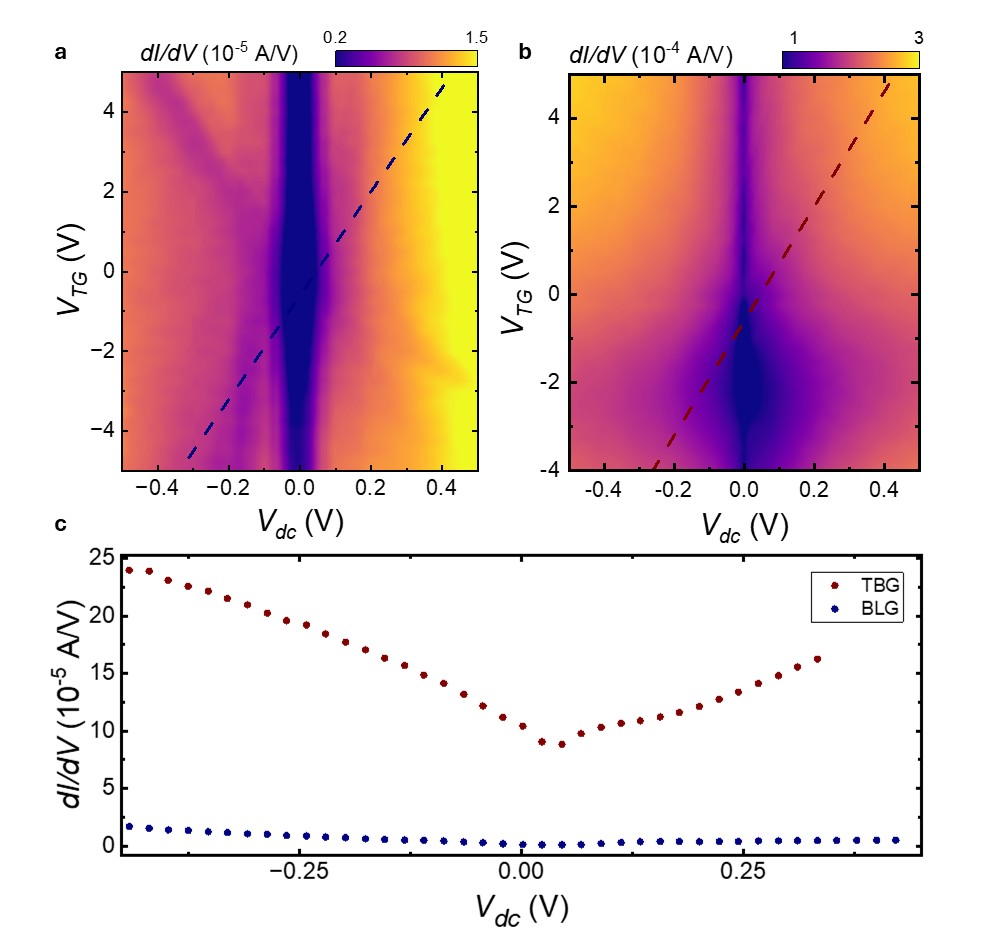}
    \captionsetup{justification=raggedright,singlelinecheck=false}
    {\textbf{Fig.~S4.~Diagonal line cuts from $V_{DC}-V_{TG}$ contour plot.}~\textbf{a.}~and~\textbf{b.}~Contour plots of tunnel conductance ($dI/dV$) as a function of $V_{dc}$ and $V_{TG}$ for BLG and tBLG, respectively (also shown in Fig.~1f~and~2c). The blue and maroon dashed lines represent the diagonal paths. ~\textbf{c.}~$dI/dV$ line cuts along the dashed lines from Fig.~S4a~and~S4b respectively illustrating enhanced value and increasing trend for the tBLG case in contrast to almost flat behaviour in BLG.}
\end{figure*}

\begin{figure*}
    \centering
    \includegraphics[width=1.1\linewidth]{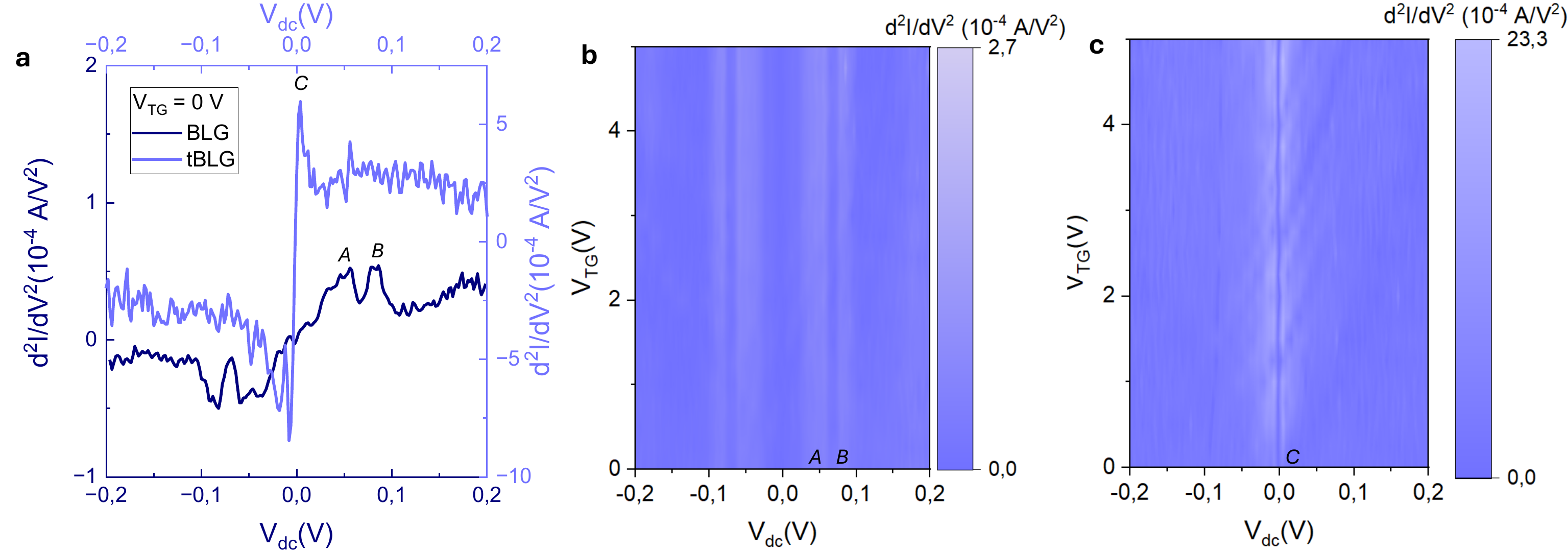}
    \captionsetup{justification=raggedright,singlelinecheck=false}{\textbf{Fig.~S5.~ Calculated $d^{2}I/dV^2$ plots for BLG and tBLG highlighting the phonon contributions}~\textbf{a.}~Line plots of $d^{2}I/dV^2$ as a function of $V_{dc}$ for BLG and tBLG at $V_{TG}=0$ V. The prominent phonons peaks are denoted by A ($\approx$ 54 meV) and B ($\approx$ 82 meV) in BLG and C ($\approx$ 8 meV) in tBLG. The sharp rise in $d^{2}I/dV^2$ around C, suggests the dominant contribution from the low energy LBM mode for tBLG. 
    \textbf{b.} and \textbf{c.}  show $d^{2}I/dV^2$ as a function of $V_{dc}$ and $V_{TG}$ for BLG and tBLG, respectively. The negligible variation in the signatures with $V_{TG}$, results in faint blue vertical lines symetrical across zero bias, indicating that these signatures likely arise from phonons.}
    \label{fig:S5}
    
\end{figure*}

\subsection{Comparison of tunneling in various metal-insulator-graphitic tunnel junctions}

 {This work discusses enhanced tunneling in twisted bilayer graphene-based tunnel junctions, and compares the data with the suppressed tunneling seen in Bernal bilayer graphene. The findings of this study builds on the literature available for metal- insulator-graphitic tunnel junctions, and therefore, in the table below, we present a comprehensive report of zero bias gaps seen across different material systems. The metals used range from Cr/Au to scanning tunneling microscopy (STM) tips made of Pt/Ir or W, while the graphitic tunnel electrodes range from single layer graphene (SLG), Bernal bilayer graphene (BLG) to graphite. The table compiles both STM and planar tunneling studies, and shows that irrespective of the finer details in the geometry as well as the measurement technique, the zero bias suppression stays strong with large gaps of $\sim 30-70$ mVs. An enhanced tunneling and a narrow zero-bias gap, as reported in our work on tBLG-based junctions has not been reported thus far, to the best of our knowledge. }
\begin{table}
    \centering
    \begin{tabular}{||c|c|c|c||}
         \hline
         Material system & Measurement scheme & Zero bias gap ($w/2$) & Reference \\
         \hline
         \hline
         Cr-Au/hBN/SLG & Planar tunneling & 35-50 mV & \cite{chandni2016signatures} \\
         \hline
         Ag/hBN/BLG & Planar tunneling & 40-50 mV & \cite{chandni2016signatures} \\
         \hline
         Cr-Au/hBN/graphite & Planar tunneling & 30-40 mV & \cite{chandni2016signatures} \\
         \hline
         Au/hBN/SLG & Planar tunneling & 50-70 mV & \cite{davenport2019probing} \\
         \hline
         Au/hBN/graphene & Planar tunneling & 50 mV & \cite{PhysRevB.85.073405} \\
         \hline
         W/Air/SLG on SiC & STM & 65 mV & \cite{Zhang2008} \\
         \hline
         W/Air/SLG on SiC & STM & 50 mV & \cite{brar2007scanning} \\
         \hline
         Pt-Ir/Air/SLG on SiC & STM & 50 mV & \cite{brar2007scanning} \\
         \hline
         Ag, Pt-Ir, W/Air/HOPG & STM & 65 mV & \cite{PhysRevB.102.115410}\\
         \hline
         W/Air/SLG-Ir & STM & 50 mV & \cite{Halle2018} \\
         \hline
         Pt-Ir/Air/SLG & STM & 65 mV & \cite{PhysRevB.100.075435} \\
         \hline
         Au/WSe2/BLG & Planar tunneling & 30-50 mV & This work \\
         \hline
         Au/WSe2/tBLG & Planar tunneling & 5-10 mV & This work \\
         \hline
    \end{tabular}
    \caption{Comparison between zero-bias gaps seen for various metal-insulator-graphitic tunnel junctions}
    \label{tab:my_label}
\end{table}

\subsection{Elastic and inelastic tunneling current}
 {We first define the combined Hamiltonian of the top and bottom electrodes. In our study, Au and tBLG are the bottom and top electrodes, respectively. The Hamiltonian is written as:}
\begin{equation}
{H} = {H}_0 + {H}_t
\end{equation}
 {where \( {H}_0 \) is the interaction part of the Hamiltonian and \( {H}_t \) is the tunneling part.}
\begin{equation} 
{H}_0 = \sum_{{k}_B} \epsilon_{{k}_B}^{B} c_{{k}_B}^{\dagger} c_{{k}_B} + \sum_{{k}_T} \epsilon_{{k}_T}^{T} c_{{k}_T}^{\dagger} c_{{k}_T} 
+ \sum_{{q},\nu} \omega_{{q}\nu} b_{{q}\nu}^{\dagger} b_{{q}\nu}+ \sum_{{k}_T, {q}, \nu} g_{{k}_T ,{q} \nu} c_{{k}_T+{q}}^{\dagger} c_{{k}_T} ( b_{{q} \nu} + b_{-{q} \nu}^{\dagger} )
\end{equation}
 {where $c_{{k}_{T}}$ and $c_{{k}_{B}}$ are the electron annihilation operators of the top and bottom electrodes, respectively, and $\epsilon_{{k}_{T}}$ and $\epsilon_{{k}_{B}}$ are the corresponding energy dispersions. $b_{{q}\nu}$ is the phonon annihilation operator for the phonon momentum, ${q}$ and the mode index, $\nu$ with the phonon dispersion $\omega_{{q}\nu}$. We take $\hbar=1$. $g_{{k}_T ,{q} \nu}$ are the electron-phonon coupling matrix elements. As Au's phonon DOS is featureless in the low energy region ({\em i.e.} it does not have any sharp peaks) and its electron-phonon coupling is expected to be small compared to graphene, it is reasonable to ignore the phonon and electron-phonon (el-ph) terms in the Hamiltonian for the bottom electrode (Au). This is particularly true at low temperatures (as all the measurements were performed below 4K). We focus on low-frequency phonon DOS of the top electrode (tBLG) which shows sharp peaks corresponding to van Hove singularities arising from layer breathing modes near the $\Gamma$ point. The tunneling Hamiltonian is written as:}
\begin{equation} 
{H}_t = \sum_{{k}_T, {k}_B} M_{{k}_T, {k}_B} c_{{k}_T}^{\dagger} c_{{k}_B} + h.c.
\end{equation}
 {Here $M_{{k}_T ,{k}_B}$ are the tunneling matrix elements, which can be expanded into elastic and inelastic parts as:}
\begin{equation}  
M_{{k}_T ,{k}_B} = M_{{k}_T, {k}_B}^{e} + \sum_{{q},\nu} M_{{k}_T ,{k}_B, {q} \nu}^{i} g_{{k}_T ,{q} \nu} \phi_{{q} \nu} + {O}(\phi_{{q} \nu}^2)
\end{equation}
 {where $\phi_{{q} \nu} = b_{{q} \nu} + b_{-{q} \nu}^{\dagger}$}

\subsubsection{Elastic tunneling current}
 {At a bias voltage \( V \), the total elastic tunneling current, using Fermi's Golden Rule, can be written as:}
\begin{equation}  
I^e (V) = 4 \pi e \sum_{{k}_T, {k}_B} |M_{{k}_T, {k}_B}^{e}|^2 f(\epsilon_{{k}_T}^{T}) (1 - f(\epsilon_{{k}_B}^{B}))
\delta(\epsilon_{{k}_T}^{T} - \epsilon_{{k}_B}^{B} - eV)
\end{equation}
  {where $f(\epsilon)$ is the Fermi occupation.}
\begin{equation} 
= 4 \pi e  \int dE \int dE' \sum_{{k}_T, {k}_B} |M_{{k}_T, {k}_B}^{e}|^2 f(E) (1 - f(E'))
\delta(E - \epsilon_{{k}_T}^{T}) \delta(E' - \epsilon_{{k}_B}^{B}) \delta(E - E' - eV)
\end{equation}
 {Defining,}
\begin{equation}  
A_{{k}_T}^{T}(E) = \delta(E - \epsilon_{{k}_T}^{T})
\end{equation}
\begin{equation}  
A_{{k}_B}^{B}(E') = \delta(E' - \epsilon_{{k}_B}^{B})
\end{equation}
 {we get,}
\begin{equation} 
I^e (V) = 4 \pi e \int dE \int dE' \sum_{{k}_T, {k}_B} |M_{{k}_T ,{k}_B}^{e}|^2  f(E) (1 - f(E'))
A_{{k}_T}^{T}(E) A_{{k}_B}^{B}(E') \delta(E - E' - eV)
\end{equation}
 {Integrating over \( E' \),}
\begin{equation} 
I^e (V) = 4 \pi e \int dE \sum_{{k}_T, {k}_B} |M_{{k}_T, {k}_B}^{e}|^2 f(E) (1 - f(E - eV)) 
A_{{k}_T}^{T}(E) A_{{k}_B}^{B}(E - eV)
\end{equation}
\begin{equation}  
M_{{k}_T, {k}_B}^{e} \approx |M^e| \delta ({k}_T - {k}_B)
\end{equation}
 {which is the momentum conservation criteria and we assume tunneling to be constant. The momentum summation results in a joint density of states for the top and bottom electrodes. We approximate this joint density of states as a product of the individual densities of states, while retaining the momentum conservation condition.}
\begin{equation}  
I^e (V) \propto \int |M^e|^2 \delta ({k}_T - {k}_B) f(E) (1 - f(E - eV)) 
\rho_T(E) \rho_B (E - eV) dE
\end{equation}
 {where,}
\begin{equation}  
\rho_T(E) = \sum_{{k}_T} A_{{k}_T}^{T} (E)
\end{equation}
\begin{equation} 
\rho_B(E - eV) = \sum_{{k}_B} A_{{k}_B}^{B} (E - eV)
\end{equation}

\subsubsection{Inelastic tunneling current}
 {Using the inelastic part of tunneling Hamiltonian, and the Fermi’s Golden Rule, the inelastic tunneling current can be written as:}
\begin{equation}
  \begin{split}
I^{i} (V) &= 4\pi e \sum_{{k}_T, {k}_B, {q}\nu} \left| M_{{k}_T, {k}_B, {q}\nu}^{i} g_{{k}_T, {q}\nu} \right|^2
 \Bigg[ \Big( f(\epsilon_{{k}_B}^{B}) n^{ph} (\omega_{{q}\nu})[1-f(\epsilon_{{k}_T}^{T})] - f(\epsilon_{{k}_T}^{T}) [1 + n^{ph} (\omega_{{q}\nu})]
 \\
 &[1 - f(\epsilon_{{k}_B}^{B})]\Big) \delta (\epsilon_{{k}_T}^{T} - \epsilon_{{k}_B}^{B} - \omega_{{q}\nu} + eV)+ \Big(f(\epsilon_{{k}_B}^{B}) [1 + n^{ph} (\omega_{{q}\nu})] [1 - f(\epsilon_{{k}_T}^{T})] - f(\epsilon_{{k}_T}^{T}) \\
 &n^{ph} (\omega_{{q}\nu})[1 - f(\epsilon_{{k}_B}^{B})] \Big) \delta (\epsilon_{{k}_T}^{T} - \epsilon_{{k}_B}^{B} + \omega_{{q}\nu} + eV)
\Bigg]
\end{split}
\end{equation}
  {where $n^{ph} (\omega)$ is the Bose occupation factor.}
\begin{equation}  
\begin{split}
I^{i} (V) &= 4\pi e \iiint d\omega dE dE' \sum_{{k}_T, {k}_B, {q}\nu} \left| M_{{k}_T ,{k}_B, {q}\nu}^{i} g_{{k}_T, {q}\nu} \right|^2\Bigg[ A_{{q}\nu}^{ph} (\omega) A_{{k}_T}^{T} (E) A_{{k}_B}^{B} (E') 
\\
&\Big( f(E') n^{ph} (\omega) [1 - f(E)] - f(E) [1 + n^{ph} (\omega)] [1 - f(E')] \Big)\delta (E - E' - \omega + eV) \\
& A_{{q}\nu}^{ph} (\omega) A_{{k}_T}^{T} (E) A_{{k}_B}^{B} (E') \Big( f(E') [1+n^{ph} (\omega)][1 - f(E)] - f(E) n^{ph} (\omega) [1 - f(E')] \Big) \\
&\delta (E - E' + \omega + eV)\Bigg]
\end{split}
\end{equation}
 {where $A_{{q}\nu}^{ph}(\omega) = \delta(\omega - \omega_{{q}\nu})$, $A_{{k}_T}^{T}(E)$ and $A_{{k}_B}^{T}(E')$ are as defined before.}
\begin{equation}
  \begin{split}
I^{i} (V) &= 4\pi e \iint d\omega dE \sum_{{k}_T, {k}_B, {q}\nu} \left| M_{{k}_T, {k}_B, {q}\nu}^{i} g_{{k}_T, {q}\nu} \right|^2\Bigg[ A_{{q}\nu}^{ph} (\omega) A_{{k}_T}^{T} (E) A_{{k}_B}^{B} (E - \omega + eV) \\
&\Big( f(E - \omega + eV) n^{ph} (\omega) [1 - f(E)] - f(E) [1 + n^{ph} (\omega)] [1 - f(E - \omega + eV)] \Big) \\
&+ A_{{q}\nu}^{ph} (\omega) A_{{k}_T}^{T} (E) A_{{k}_B}^{B} (E + \omega + eV)\Big( f(E + \omega + eV) [1 + n^{ph} (\omega)] [1 - f(E)] \\ 
&\Big( f(E + \omega + eV) [1 + n^{ph} (\omega)] [1 - f(E)]- f(E) n^{ph} (\omega) [1 - f(E + \omega + eV)] \Big) \Bigg] 
\end{split}
\end{equation}
 
{In the low-frequency range, where the phonon DOS exhibits a sharp peak associated with the layer breathing modes, both the electron-phonon coupling strength and inelastic tunneling matrix elements are not expected to vary significantly with momentum. We therefore approximate them as constants.}
\begin{equation}
  M_{{k}_T, {k}_B, {q}\nu}^{i} \approx |M^i| \delta ({k}_T - {k}_B \pm {q})
\end{equation}
\begin{equation}
  g_{{k}_T, {q}\nu} \approx g
\end{equation}
 
{Here, the phonon spectral function will now be weighted by the average electron-phonon coupling. Analogous to elastic tunneling current, we approximate the joint density of states as a product of the individual densities of states in this case too, while retaining the momentum conservation condition. $I^{i}(V)$ will depend on the following quantities for all the terms of phonon emission and absorption processes, upto the corresponding occupation factors, as follows:}
\begin{equation}
  I^{i} (V) \propto |M^i|^2 |g|^2 
\iint \delta ({k}_T - {k}_B \pm {q}) \rho^{ph} (\omega) \rho_T (E) \rho_B (E \pm \omega + eV) \, d\omega dE
\end{equation}
 {where $\rho^{ph}(\omega) = \sum_{{q}\nu}A_{{q}\nu}^{ph}(\omega)$, $\rho_{T}(E)$ and $\rho_{B}(E\pm \omega+eV)$ are as defined before.}

\subsection{Discussion on other possibilities}
 {We have attributed the enhanced tunneling observed and the narrow zero bias gap observed in tBLG to the layer breathing modes, and the relaxed momentum conservation in tBLG due to the reduced moir\'{e} Brillouin zone. In the following we discuss few other possibilities for the signatures observed, and highlight why LBM phonon-mediated tunneling is the most plausible scenario.}


\subsubsection{ZA phonon modes}

 {In the energy range we are investigating, earlier studies \cite{doi:10.1021/acs.jpclett.1c01802} have shown for BLG on Ru (0001), the ZA phonon modes appear in the $d^2I/dV^2$ spectrum at finite energies of $\sim$ 13 meV instead of commencing from zero at the $\Gamma$-point, as also supported by theory. Naturally, this indicates the possibility of a similar mechanism in our system as well. But our case is similar to the free graphene case, and our calculations do not show that the ZA phonons have a finite energy at the $\Gamma$-point. Instead, around the same energy scales, the layer breathing optical mode is observed at the $\Gamma$-point agreeing with our experimental observations. We also do not observe any significant effect of the ZA mode phonon in the BLG case, and therefore conclude that the free-standing approximation is valid.}

\subsubsection{Backfolding effects}

 {As discussed in Ref. \cite{endlich2014moire} phonons in moir\'{e} systems might exhibit the effects of backfolding, resulting in replica phonon peaks in tunneling, shifted by the moir\'{e}  reciprocal lattice vector ${q} \pm {G}_{K/M}$, along the $\Gamma-K$ or $\Gamma-M$ directions. We do not find any evidence for such replica peaks in our current data. We suspect this is because the lattice periodicity of $7.235$ nm (corresponding to $\theta \approx 2^\circ$), gives $|{G}_K|\approx 0.86$ nm$^{-1}$ and $|{G}_M| \approx 0.5$ nm$^{-1}$ , which are significantly smaller than the graphene on  Ir(111) case. Therefore, we speculate that the signatures are either too weak to be observed, or irrelevant for the present study.}

\subsubsection{Three terminal model}

{Some of the previous  studies have followed a three terminal model \cite{Halle2018, kroger2020local} where, in addition to the two conventional tunneling terminals as source and substrate, a third terminal is introduced which can collect electron propagation in graphene alone. We conclude that such as a model is not directly applicable to our experiments. Our devices consist of two terminals, namely (1) Au electrode and (2) tBLG or BLG, which are separated by a few layers of WSe$_2$. We assume that the bands in tBLG are hybridized at the low-twist angle regime and is therefore a homogeneous system Hence, the presence of a third terminal is not compatible with this device geometry.}

\subsubsection{Effects of the moir\'{e} potential}

{In this work, we have particularly worked with a non-magic angle sample. Therefore, strong correlation effects are not expected and our transport data do not show evidences for correlated insulators, superconductivity etc, either. Therefore, we suspect that the tunneling features are not directly related to the moir\'{e} bands, in terms of strong correlations. As discussed in the main text, the moir\'{e} pattern does enhance the  geometric overlaps for the Fermi surfaces in case of tBLG samples. While the complex nature of moir\'{e} flat bands can introduce finer modifications in the tunneling response, further studies particularly near the magic angle are required to explore these effects in detail.}

\end{document}